\documentclass{article}
\usepackage{mathtools}
\usepackage{enumitem}
\usepackage{spconf,amsmath,graphicx}
\usepackage{subcaption} 
\usepackage{amssymb}
\usepackage{epsfig}
\usepackage{caption3}
\usepackage{acronym}
\usepackage{fancyhdr}
\usepackage{color}
\usepackage{verbatim}
\usepackage{listings}
\usepackage{graphicx}
\usepackage{cite}
\usepackage{caption}
\usepackage{hyperref}
\usepackage{subcaption}
\usepackage{multirow}

\def\a{{\mathbf a}}

\def\w{{\mathbf w}}
\def\x{{\mathbf x}}
\def\y{{\mathbf y}}

\def\0{{\mathbf 0}}
\def\1{{\mathbf 1}}

\def\A{{\mathbf A}}

\def\E{{\mathbf E}}

\def\I{{\mathbf I}}

\def\X{{\mathbf X}}
\def\Y{{\mathbf Y}}
\def\W{{\mathbf W}}

\def\bgamma{{\boldsymbol{\gamma}}}

\def\bSigma{{\boldsymbol{\Sigma}}}

\def\cC{{\cal C}}

\def\N{{\cal N}}

\def\C{{\mathbb C}}

\title{PARAMETRIC MODELS FOR DOA TRAJECTORY LOCALIZATION}

\name{Ruchi Pandey and 
Santosh Nannuru}
\address{IIIT Hyderabad, SPCRC, Hyderabad, India
	}
\begin{document}
%
\maketitle
\begin{abstract}
Directions of arrival (DOA) estimation or localization of sources is an important problem in many applications for which numerous algorithms have been proposed. Most localization methods use block-level processing that combines multiple data snapshots to estimate DOA within a block. The DOAs are assumed to be constant within the block duration. However, these assumptions are often violated due to source motion. In this paper, we propose a signal model that captures the linear variations in DOA within a block. We applied conventional beamforming (CBF) algorithm to this model to estimate linear DOA trajectories. Further, we formulate the proposed signal model as a block sparse model and subsequently derive sparse Bayesian learning (SBL) algorithm. Our simulation results show that this linear parametric DOA model and corresponding algorithms capture the DOA trajectories for moving sources more accurately than traditional signal models and methods.
\end{abstract}


\section{Introduction}
Source directions of arrival (DOA) estimation is a crucial task in many applications such as channel estimation \cite{prasad2014joint}, radar \cite{herman2009high}, acoustic array processing \cite{gerstoft2015multiple}, smart devices \cite{wan2016application}, and hearing aids \cite{farmani2015maximum}. Along with conventional beamforming (CBF) \cite{vantrees2002} and multiple signal classification (MUSIC) \cite{music1986}, compressive sensing based sparse reconstruction \cite{Compressivesensing} and sparse Bayesian learning (SBL) \cite{tipping2001,gerstoft2016,pandey2021sparse} are some popular methods for DOA estimation. Almost all localization algorithms use block-level processing where a block consists of multiple data snapshots which are processed together to estimate the source DOA. Block-level processing is important, especially in low signal-to-noise ratio (SNR) scenarios, to obtain reliable DOA estimates. In case of moving sources, the DOA change within the block duration and it is required that the sources are tracked over time. This is commonly done by processing overlapping blocks and applying tracking filters such as Kalman filter~\cite{kalman1960new} or particle filter~\cite{pf2002tutorial} over block DOA estimates to obtain source motion trajectories. Recently some works have addressed the problem of DOA trajectory estimation using Bayesian analysis  \cite{park2021sequential} and neural networks \cite{opochinsky2021deep,diaz2020robust}. 

In this work, we propose a signal model which incorporates source motion using parametric trajectories and accounts for linear DOA motion within the block duration. This provides better DOA estimates compared to models assuming fixed DOA. It can also be extended to other parametric models albeit with a further increase in processing complexity (in this paper we focus on the linear motion which requires two parameters per source). Parametric trajectories have the potential to eliminate the need for tracking filters by implicitly performing both localization and tracking. We refer to this as trajectory localization (TL).
%
~In this paper we, \\
{\bf{(a)}} introduce a signal model which incorporates linear DOA motion within a block; \\
{\bf{(b)}} develop an extension of CBF, called TL-CBF, to perform parametric trajectory estimation; \\
{\bf{(c)}} reformulate model {\bf{(a)}} in a sparse signal framework and develop TL-SBL algorithm for trajectory localization.

\section{Signal Model}
\label{sec:proposed model}

\subsection{Static DOA}
The measurement $\y \in \C^{N}$ recorded by an $N$-sensor uniform linear array (ULA) when $K$ sources are present is
\begin{align}
    \y &= \sum_{k=1}^{K}{\a(\theta_{k}) s_{k}} + \w = \A \x + \w \,,
\end{align} 

where $\a(\theta_{k}) = \textbf{a}_{k} = [1,e^{j2\pi \frac{d}{\lambda}\sin\theta_k},\hdots, e^{j2\pi (N-1)\frac{d}{\lambda}\sin \theta_k}]$ 
is the steering vector corresponding to the source in direction $\theta_{k}$, steering vector matrix $\A = [\a_{1} \ldots \a_{K}]$, source amplitude vector $\x = [s_1, \ldots, s_K]$, and $\w \in \C^{N}$ is the additive noise. Here, $\lambda$ is the wavelength of the narrowband sources and $d$ is the separation between the sensors in the ULA. When multiple observations are available, assuming the source DOA are unchanging, the multiple measurement vector (MMV) model~\cite{wipf2007,gerstoft2016} is given by
\begin{align}
 \textbf{Y} = \textbf{AX} + \textbf{W}
    = [\A\x_{1} \ldots \A\x_{L}] + \textbf{W} \,,
  \label{sigmodel}
\end{align}
where $\Y = [\y_1 \hdots ~\y_L] \in \C^{N\times L}$ is the $L$-snapshot measurement matrix, $\X = [\x_1 \hdots \x_L] \in \C^{K\times L}$ denotes source amplitudes across snapshots, and $\W = [\w_1 \hdots \w_L] \in \C^{N\times L}$ accounts for the additive noise across $L$ snapshots. In block-level processing, each $L$-snapshot block is used to estimate the DOA parameters $\theta_k, k = 1,\ldots,K$. 


\subsection{Linear DOA trajectory}

In most practical cases the sources are moving and the unchanging DOA assumption no longer holds. As a first order correction to this, we propose to incorporate linear DOA motion across snapshots within a block. Linearly changing DOA $\theta^{l}$ as a function of snapshot number $l$ can be captured using a two parameter $(\phi,\alpha)$ model as
\begin{align}
    \theta^{l} = \phi + \frac{l-1}{L-1} \alpha, \quad l = 1,2,\ldots,L
    \,, \label{linear_doa}
\end{align}
where $\phi$ is the DOA in the first snapshot and $\phi + \alpha$ is the DOA in the last snapshot of an $L$-snapshot block, i.e. the DOA changes by $\frac{\alpha}{L-1}$ across each snapshot. The parameter pair $(\phi,\alpha)$ captures DOA motion within a block. Define $\tilde{\A}(\phi,\alpha) \in \C^{N \times L}$ to be the matrix of all steering vectors as the DOA changes with linear model parameters $(\phi,\alpha)$, i.e. $\tilde{\A}(\phi,\alpha) = [\a(\theta^{1}) \ldots \a(\theta^{L}))]$. The multiple measurement vector model accounting for linear DOA motion is 
\begin{align}
    \Y &= \sum_{k=1}^{K}{\tilde{\A}(\phi_{k},\alpha_{k})} \tilde{\X}_{k} + \W 
    = \sum_{k=1}^{K}{\tilde{\A}_{k} \tilde{\X}_{k}} + \W \,, \\
    \Y &= \tilde{\A} \tilde{\X} + \textbf{W} \,,
    \label{newsigmodel}
\end{align}
where $\tilde{\X}_{k} = \text{diag}(\x^{k})$, $\x^{k} = [s^{1}_{k} \ldots s^{L}_{k}]^{T}$ is the vector of $L$ amplitudes of the $k$th source, $\tilde{\A} = [\tilde{\A}_{1} \hdots  \tilde{\A}_{K}] \in \mathbb{C}^{N \times KL}$, and $\tilde{\X} = [\tilde{\X}_{1} \hdots  \tilde{\X}_{K}]^{T} \in \mathbb{C}^{KL\times L}$. Here $\tilde{\X}_{k} \in \C^{L \times L}$ is a diagonal matrix with $k$th source amplitudes across $L$ snapshots on the diagonal.

Equation \eqref{newsigmodel} is the MMV model which accounts for linear DOA motion.
In linear trajectory localization, our goal is to estimate the parameter pairs $(\theta_{k},\alpha_{k})$ for all the  sources and hence obtain their linear trajectory estimates within a block. It is straightforward to extend the linear motion in  \eqref{linear_doa} to higher order polynomials or any other parametric trajectory. Computational requirements of the DOA estimation algorithms will grow with the number of parameters in the model. In this paper, we demonstrate the idea of trajectory localization by focusing on linear trajectories.

\subsection{Conventional beamforming}
\label{sec:CBF-M}
We propose a modification of the conventional beamforming (CBF) \cite{compressivebeamforming} algorithm for the signal model presented above. We refer to it as trajectory localization CBF, i.e. TL-CBF. In CBF, the angular power spectrum is computed at a predefined grid by evaluating the correlation between the observations and the steering vectors. The peaks of this angular power spectrum provide DOA estimates.


Extending this notion, the TL-CBF power spectrum for $L$ snapshots is computed as
\begin{align}
    P_{TL-CBF}(\phi,\alpha) =  \frac{1}{L} \sum_{l=1}^{L} |\a^H_l(\phi,\alpha)~\y_l|^2
    \,, \label{cbf_spec_final}
\end{align}
where the power spectrum $P_{TL-CBF}(\phi,\alpha)$ is now two dimensional. The locations of the peaks in this spectrum provide DOA trajectory estimates. Note that \eqref{cbf_spec_final} is different from the  traditional MMV CBF as each steering vector now incorporates information about potentially changing DOA.

\section{Sparse Signal Model}
\label{sec:SBL-M}
In this section, model in \eqref{newsigmodel} is reformulated as a sparse signal model allowing us to apply sparse signal processing algorithms for DOA estimation. For sparse formulation, consider a finely sampled grid in $(\phi,\alpha)$ space. Let the uniformly sampled points in this rectangular space be denoted by $\{ (\phi_{1},\alpha_{1}), \ldots, (\phi_{1},\alpha_{M_2}), \ldots, (\phi_{M_1},\alpha_{M_2}) \}$. A sparse model for \eqref{newsigmodel} can be written as
\begin{align}
    \Y &= \sum_{m_1=1}^{M_1} \sum_{m_2=1}^{M_2} {\tilde{\A}(\phi_{m_1},\alpha_{m_2})} \tilde{\X}_{m_1 m_2} + \W \,, \\
    &= \sum_{m_1=1}^{M_1} \sum_{m_2=1}^{M_2} {\tilde{\A}_{m_1,m_2} \tilde{\X}_{m_1,m_2}} + \W \,, \\
    &= \tilde{\A}_s ~\tilde{\X}_{s} + \textbf{W} \,,
    \label{sparsesigmodel}
\end{align}
where $\tilde{\A}_{m_1,m_2} = \tilde{\A}(\phi_{m_1},\alpha_{m_2})$ and $\tilde{\X}_{m_1,m_2}$ are the changing DOA steering vector matrix and source amplitude matrix for the source at $(\phi_{m_1},\alpha_{m_2})$. Among all the potential $M_1 M_2$ sources, only a few (K) are present in a given block. This sparsity is modeled by matrices $\tilde{\X}_{m_1,m_2}$, only K of which are non-zero. Here we assume that the true sources lie on the grid. For compact expression we define, $\tilde{\A}_s = [\tilde{\A}_{1,1} \hdots  \tilde{\A}_{M_1,M_2}] \in \mathbb{C}^{N \times M_1 M_2 L}$, and $\tilde{\X}_s = [\tilde{\X}_{1,1} \hdots  \tilde{\X}_{M_1,M_2}]^{T} \in \mathbb{C}^{M_1 M_2 L\times L}$.

The above MMV model can be equivalently written as a single measurement model (SMV) \cite{zhang2011sparse,zhang2013extension} by vectorizing the observation matrix $\Y$ and appropriately changing the terms on right hand side. Performing a column-wise vectorization operation on $\Y$ we get
\begin{align}
\text{vec}(\Y) &= \y_{v} = \tilde{\A}_{v} \tilde{\x}_{v} + \w_{v} \,,
\label{smvmodel} \\
\tilde{\A}_{v} &= [\I_{L} \otimes \tilde{\A}_{1,1}, \ldots, \I_{L} \otimes \tilde{\A}_{M_1,M_2}] \,, \\
\tilde{\x}_{v} &= [\text{diag}(\tilde{\X}_{1,1})^{T}, \ldots, \text{diag}(\tilde{\X}_{M_1,M_2})^{T}]^{T}\,, \\
\w_{v} &= \text{vec}(\W) \,,
\end{align}
where $\I_{L} \otimes \tilde{\A}_{m_1,m_2} \in \C^{N L \times L}$ is the column-wise Kronecker product (Khatri–Rao product) of $\I_{L}$ and $\tilde{\A}_{m_1,m_2}$, and $\I_{L}$ is the $L \times L$ identity matrix. Here the diag($\cdot$) operation on a square matrix returns the diagonal of the matrix as a column vector. The sparsity structure of matrix $\tilde{\X}_s$ is translated into block sparse structure of the vector $\tilde{\x}_{v}$. In next section we adapt sparse Bayesian learning (SBL) \cite{tipping2001} algorithm to signal model \eqref{smvmodel} giving trajectory localization SBL i.e. TL-SBL.

\subsection{Sparse Bayesian Learning}
Sparse Bayesian learning is a compressive sensing method to solve parameter estimation problems \cite{tipping2001,wipf2003bayesian,SBLprespective,wipf2007}. SBL has been investigated multiple times for DOA estimation \cite{gemba2017multi,gemba2019robust,multifreq2019sparse,nannuru2021sparse,MLEefficient}. Here we derive the TL-SBL update rule following the approach in \cite{gemba2017multi,gemba2019robust,multifreq2019sparse,nannuru2021sparse}. The block sparse structure of $\tilde{\x}_{v}$ has similarities with  the static DOA MMV model \cite{zhang2011sparse,zhang2013extension}.


\textbf{Prior:} We assume that source amplitudes are i.i.d across snapshots having zero-mean complex Gaussian distribution
\begin{align}
    p(\text{diag}(\tilde{\X}_{m})) \sim \cC \N (\0,\gamma_m \I_L) \,,
    \label{Xmmv_prior}
\end{align}

where $\gamma_m$ is the variance. In this section, we use a simplified notation where the double index $(m_1,m_2)$ is replaced by a single index $m$ and correspondingly the indices $\{ (1,1),\ldots,(M_1,M_2) \}$ are renumbered as $\{ 1, 2, \ldots, M_1 M_2 \}$. We additionally assume the amplitudes are independent across sources. Thus the unknown $\tilde{\x}_{v}$ is Gaussian distributed and parametrized by the vector $\bgamma = [\gamma_1, \ldots, \gamma_{M_1 M_2}]$.


\textbf{Likelihood:} Assuming the noise to be zero-mean complex Gaussian distributed and i.i.d across sensors and snapshots, the data likelihood can be given as
\begin{align}
    p(\y_{v}|\tilde{\x}_{v};\sigma^2) = \cC \N(\y_{v}; \tilde{\A}_v \tilde{\x}_{v},\sigma^2 \I_{NL}) \,,
    \label{ysmv_likelihood}
\end{align}

where $\sigma^2$ is the noise variance. 

\textbf{Evidence:} In SBL, $\bgamma$ is estimated using evidence maximization (or type-II maximum likelihood) where evidence is
\begin{align}
    p(\y_{v};\bgamma) = \int_{\tilde{\x}_{v}} p(\y_{v}|\tilde{\x}_{v};\sigma^2) ~p(\tilde{\x}_{v};\bgamma) ~d\tilde{\x}_{v} \,.
\end{align}
Since both prior and likelihood are Gaussian, from properties of Gaussian densities, we get evidence $p(\y_{v};\bgamma)$ to be Gaussian with zero-mean and let the covariance matrix be $\bSigma_{\y_{v}}$. The log-evidence can thus be expressed as
\begin{align}
    \log p(\y_{v};\bgamma) & \propto \log |\bSigma_{\y_{v}}| - \y_{v}^H \bSigma_{\y_{v}}^{-1} \y_{v} \,,
    \label{logevidence}
    \\
    \text{where} \quad \bSigma_{\y_{v}} &= \sigma^2 \I_{NL} + \tilde{\A}_v \bSigma_0 \tilde{\A}_v^T \,, \label{datacov} \\
    \bSigma_0 &= \E(\tilde{\x}_{v} \tilde{\x}_{v}^H) \,.
\end{align}
Evidence maximization can be performed by expectation maximization (EM) algorithm  \cite{dempster1977maximum,zhang2011sparse,zhang2013extension}, but its convergence is known to be slow \cite{tipping2001,wipf2007}. Here we derive a fixed point update rule  \cite{gemba2017multi,gemba2019robust,multifreq2019sparse,nannuru2021sparse}. To obtain the TL-SBL update rule, differentiate \eqref{logevidence} with respect to $\gamma_m$, equate the derivative to zero, and rearrange the terms to give
\begin{align}
    \hat \gamma_m^{new} = \hat \gamma_m^{old} \, \frac{\y_{v}^H \bSigma_{\y_{v}} \hat{\A}_m \hat{\A}_m^H \bSigma_{\y_{v}}^{-1} \y_{v}}{\text{Tr}[\bSigma_{\y_{v}}^{-1} \hat{\A}_m \hat{\A}_m^H] } \,,
\end{align}

where $\hat{\A}_m = \I_{L} \otimes \tilde{\A}_{m}$, and $\text{Tr}[\cdot]$ denotes trace of a matrix. Note that the update for the $m$th grid point, which corresponds to $(m_1,m_2)$ pair, depends on the matrix $\tilde{\A}_{m}$ which captures all the steering vectors of the linear DOA motion through the parameters $(\phi_{m_1},\alpha_{m_2})$. Due to hierarchical probabilistic modeling of SBL, the parameter vector $\bgamma$ is sparse \cite{tipping2001}. This sparsity in $\bgamma$ is reflected as block sparsity in the source amplitude vector $\tilde{\x}_{v}$. Thus the locations of non-zero entries of $\bgamma$ signify the source DOA trajectory estimates.
Though noise variance $\sigma^2$ can also be estimated using various methods \cite{bohme1985source,gerstoft2016,multifreq2019sparse,nannuru2021sparse}, in this paper we assume noise variance to be known for simplicity.

\vspace{-0.3cm}
\begin{figure}[b]
    \centering
	\includegraphics[width=0.44\textwidth]{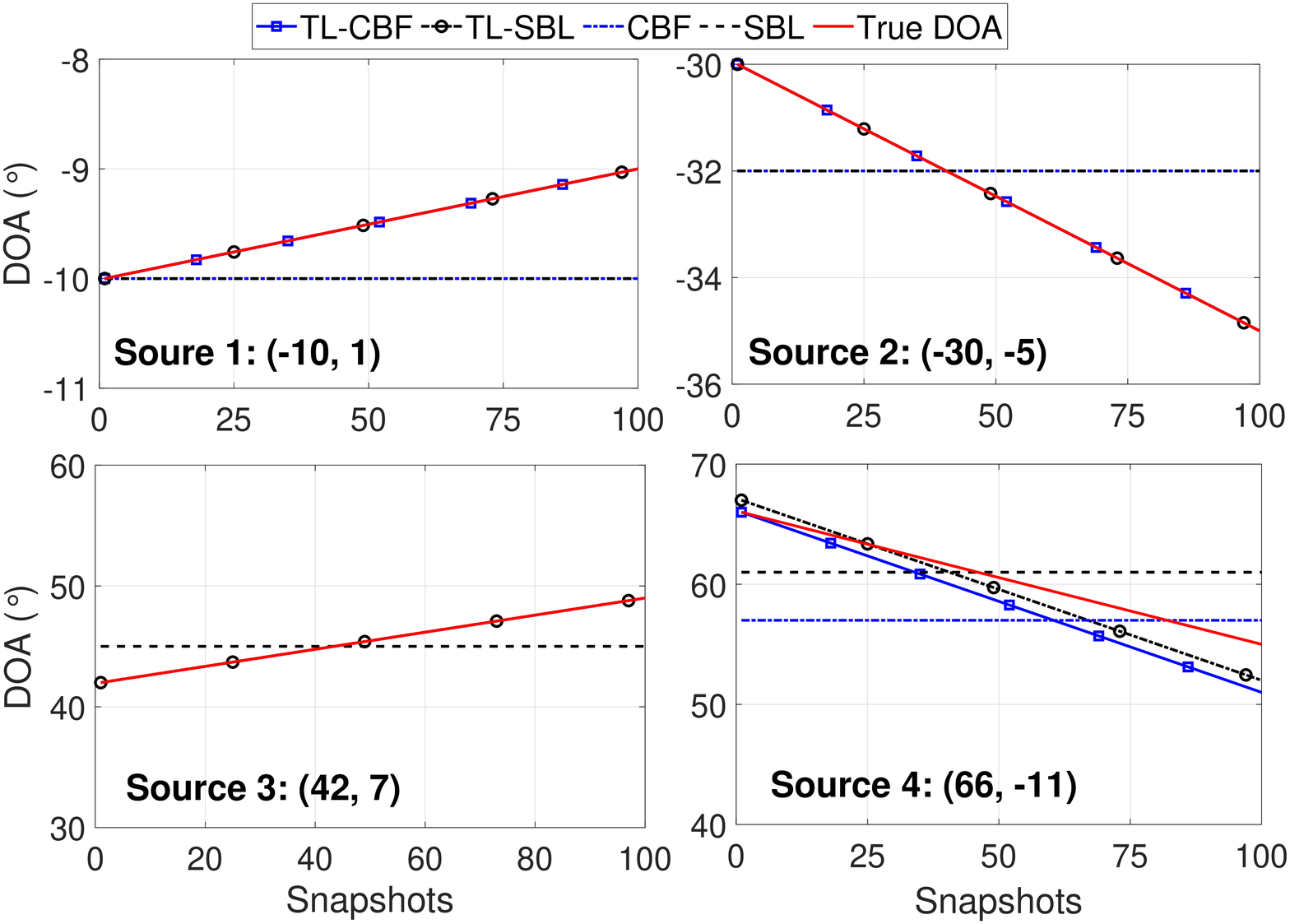}
	\caption{({\bf{Example 1}}) On-grid single block DOA estimates obtained from various algorithms for $K = 4$ sources.}
	\label{fig:ongrid_doa}
\end{figure}

\section{Simulations and results}
\label{sec:results}
We demonstrate DOA trajectory localization using the proposed signal model and algorithms on single and multiple sources. 
We compare the localization ability of TL-CBF and TL-SBL with CBF and SBL.
A uniform linear array with 10 sensors and $d = \frac{\lambda}{2}$ spacing is considered. TL-CBF and TL-SBL algorithms require a grid over the parameters $\phi$ and $\alpha$. We choose $\phi$ in the range $[-90^{\circ},90^{\circ}]$ with $1^{\circ}$ separation and $\alpha$ in the range $[-15,15]$ with $1$ unit separation. For CBF and SBL algorithms we set $\theta$ grid in the range $[-90^{\circ},90^{\circ}]$ with $1^{\circ}$ separation. The signal-to-noise ratio (SNR) is 10 dB.\\
\begin{figure*}[t!]
    \centering
	\includegraphics[width=0.44\textwidth]{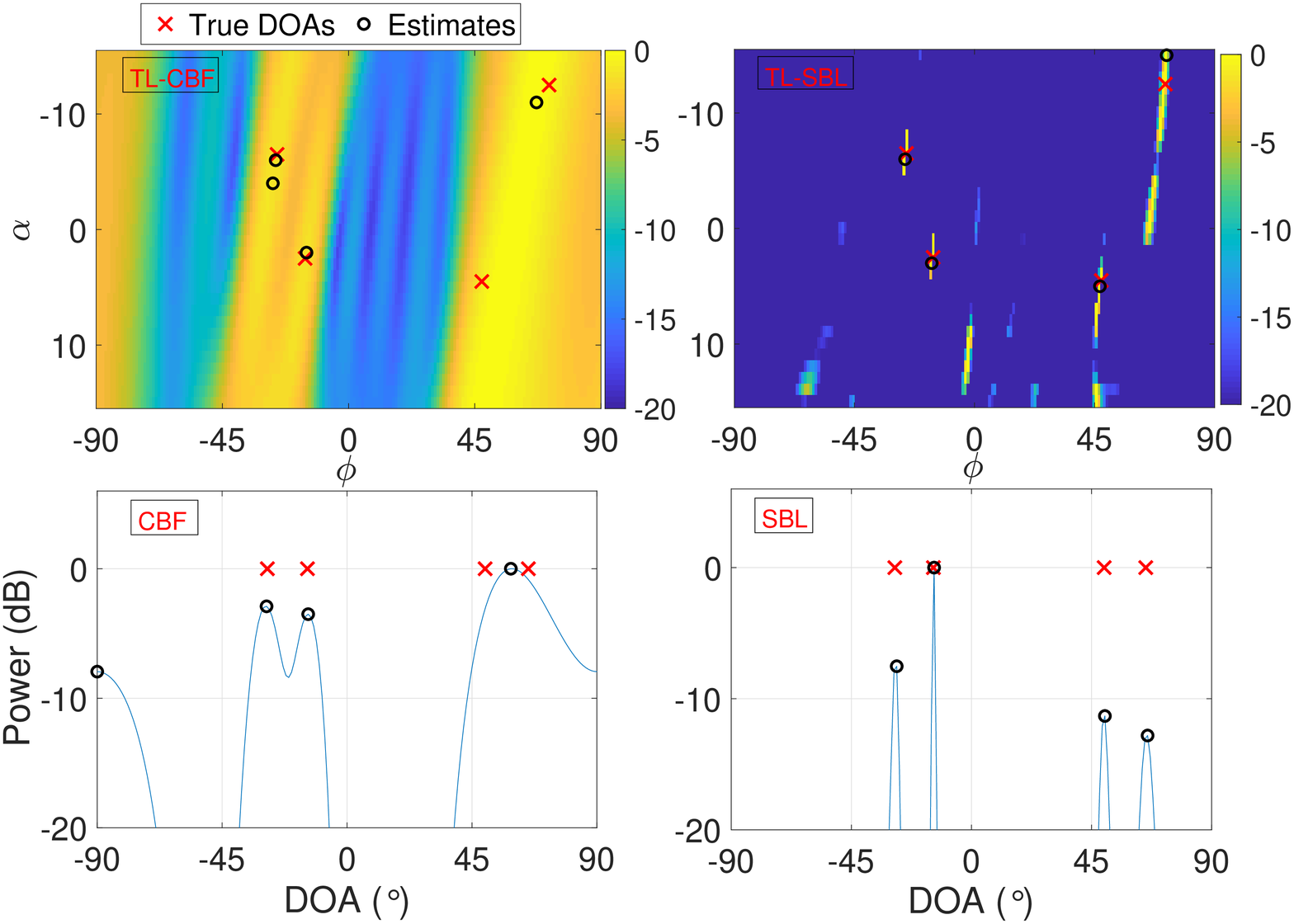}
	\includegraphics[width=0.44\textwidth]{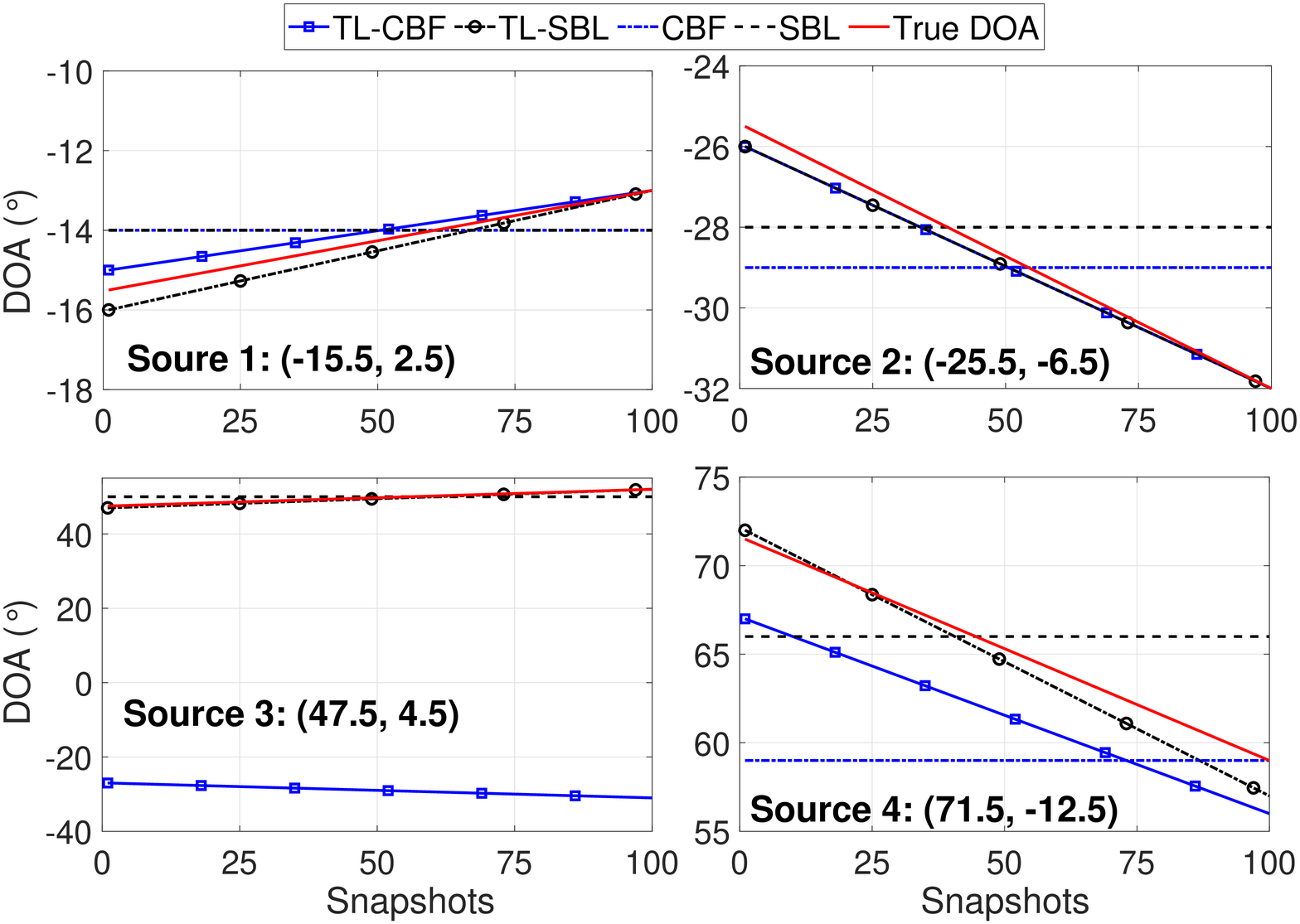}
	\caption{({\bf{Example 2}}) Power spectrum (left) of $K = 4$ off-grid sources obtained from TL-CBF, TL-SBL, CBF, and SBL algorithms at 10 dB SNR and their DOA estimates (right).}
	\label{fig:spectrum}
\end{figure*}

\begin{figure}[b!]
    \centering
	\includegraphics[width=0.44\textwidth]{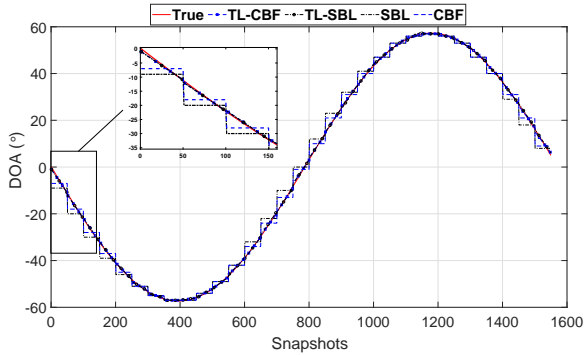}
	\caption{({\bf{Example 3}}) DOA estimates of CBF, TL-CBF, SBL, and TL-SBL for a moving source (10 dB SNR).}
	\label{fig:single_moving}
\end{figure}

{\bf{Example 1:}} We consider $K = 4$ \emph{on-grid} sources in a 100-snapshot block with linearly changing DOA. The true linear DOA parameters $(\phi,\alpha)$ for the sources are $(-10,1)$, $(-30,-5)$, $(42,7)$, and $(66,-11)$. Note that on-grid sources with $(\phi,\alpha)$ parametrization have changing DOA. The DOA estimates of all algorithms are shown in Fig.~\ref{fig:ongrid_doa}. Trajectories are obtained from estimated $(\phi,\alpha)$ parameters using \eqref{linear_doa}. 
Both TL-CBF and TL-SBL are able to estimate changing DOA within the block whereas CBF and SBL can only provide constant DOA estimates by design. Both CBF and TL-CBF suffer from poor resolution for nearby sources and some of their estimates are not visible in the region shown. Although both SBL and TL-SBL identify all four sources, TL-SBL can better adapt to linearly changing DOA.    

{\bf{Example 2:}} Here we consider $K = 4$ \emph{off-grid} sources with $(\phi,\alpha)$ parameters  $(-15.5,2.5)$,$(-25.5,-6.5)$,$(47.5,4.5)$, and $(71.5,-12.5)$ in a 100-snapshot block. The power spectrum in $(\phi,\alpha)$ domain is shown in Fig.~\ref{fig:spectrum} (left) for TL-CBF \& TL-SBL (top row) and in $\theta$ domain for CBF \& SBL (bottom row). Both SBL and TL-SBL can identify all the off-grid sources whereas CBF and TL-CBF miss a source. The corresponding DOA trajectories are shown in Fig.~\ref{fig:spectrum} (right). The TL-CBF and TL-SBL algorithms are able to find accurate on-grid approximations of the true off-grid trajectories.\\


{\bf{Example 3:}} We simulate a moving source with non-linear DOA trajectory as shown in Fig.~\ref{fig:single_moving}. The trajectory contains $31$ non-overlapping blocks of $L = 50$ snapshots each. Within each block, the DOA trajectory is approximately linear. The maximum change in DOA within any block is $11.5^\circ$. Trajectories estimated from TL-CBF and TL-SBL closely align with the true trajectory whereas CBF and SBL provide fixed DOA estimates in each block.



\begin{figure}[htb!]
    \centering
	\includegraphics[width=0.44\textwidth]{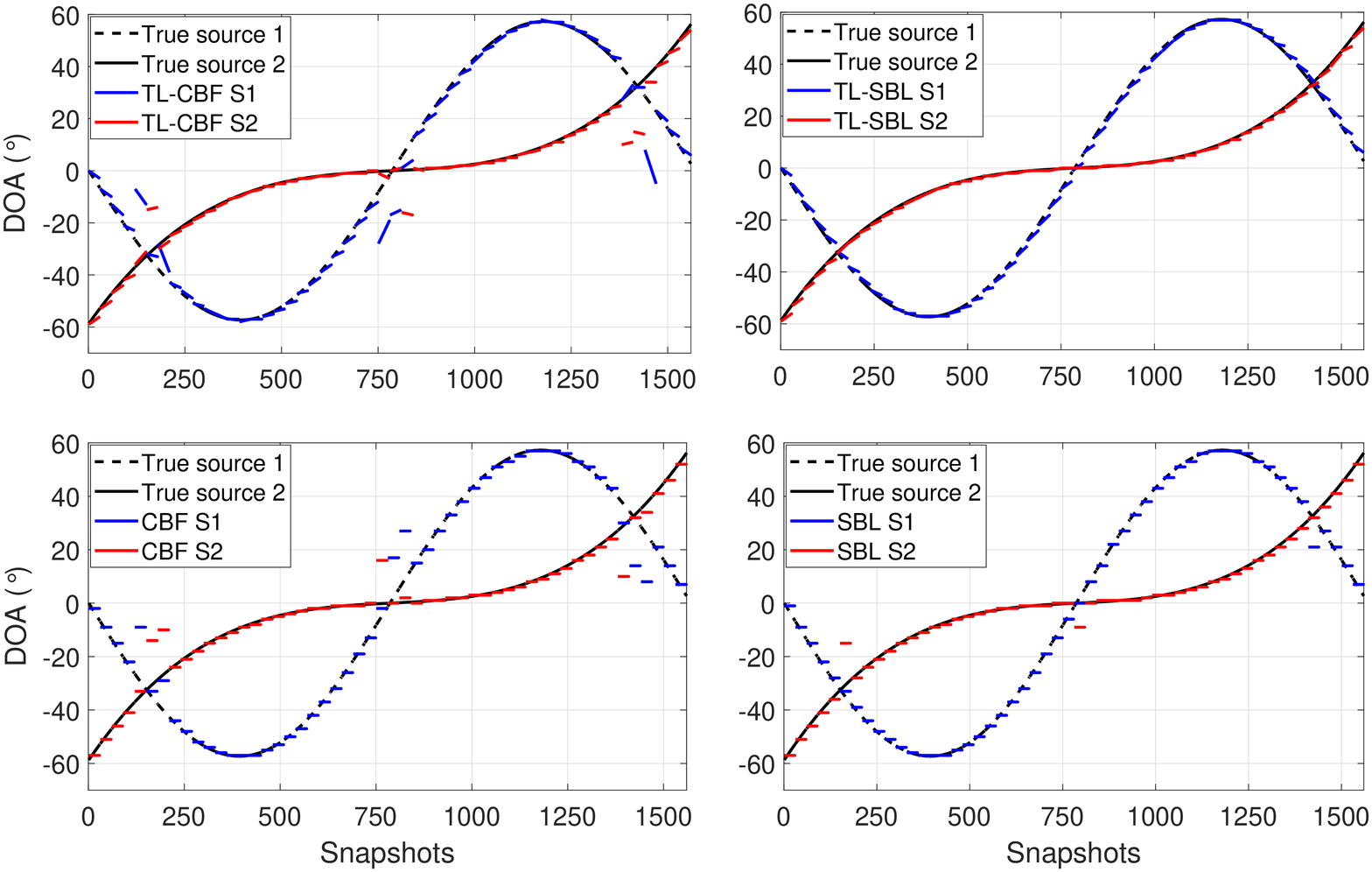}
	\caption{({\bf{Example 4}}) DOA estimates of CBF, TL-CBF, SBL, and TL-SBL for two moving sources (10 dB SNR).}
	\label{fig:two_moving}
\end{figure}

{\bf{Example 4:}} Two moving sources with non-linear DOA trajectories are simulated in Fig.~\ref{fig:two_moving} ($52$ blocks with $L = 30$ snapshots each). The estimated DOAs by TL methods provide relatively smoother trajectories. The root-mean-square DOA error for non-crossing regions are $2.98^{\circ}$, $3^{\circ}$, $2^{\circ}$, and $1.78^{\circ}$ for CBF, TL-CBF, SBL, and TL-SBL respectively.
\vspace{-0.3cm}


\section{Conclusions}
\label{sec:conclusions}
\vspace{-0.3cm}
In this work, we introduced a signal model for identifying linearly changing DOA for multi-snapshot block-level processing. 
We developed the algorithms TL-CBF and TL-SBL for the estimation of linear DOA trajectories. The analysis can be easily extended to non-linear parametric models. 
Though TL-CBF is able to estimate changing DOA, it still has the drawback of low angular resolution inherited from CBF which affects its performance in presence of multiple sources. TL-SBL improves over SBL by providing estimates of changing DOA while simultaneously having high resolution associated with compressive sensing methods.



\small
\bibliographystyle{IEEEbib}
\bibliography{strings,refs}

\begin{thebibliography}{10}

\bibitem{prasad2014joint}
R.~Prasad, C.R. Murthy, and B.D. Rao,
\newblock ``Joint approximately sparse channel estimation and data detection in
  ofdm systems using sparse {B}ayesian learning,''
\newblock {\em IEEE Trans. Signal Process.}, vol. 62, no. 14, pp. 3591--3603,
  2014.

\bibitem{herman2009high}
M.~A. Herman and T.~Strohmer,
\newblock ``High-resolution radar via compressed sensing,''
\newblock {\em IEEE Trans. Signal Process.}, vol. 57, no. 6, pp. 2275--2284,
  2009.

\bibitem{gerstoft2015multiple}
P.~Gerstoft, A.~Xenaki, and C.~F. Mecklenbr{\"a}uker,
\newblock ``Multiple and single snapshot compressive beamforming,''
\newblock {\em J. Acoust. Soc. Am.}, vol. 138, no. 4, pp. 2003--2014, 2015.

\bibitem{wan2016application}
L.~Wan, G.~Han, L.~Shu, S.~Chan, and T.~Zhu,
\newblock ``The application of {DOA} estimation approach in patient tracking
  systems with high patient density,''
\newblock {\em IEEE Trans. Ind. Electron.}, vol. 12, no. 6, pp. 2353--2364,
  2016.

\bibitem{farmani2015maximum}
M.~Farmani, M.~S. Pedersen, Z.~H. Tan, and J.~Jensen,
\newblock ``Maximum likelihood approach to “informed” sound source
  localization for hearing aid applications,''
\newblock in {\em IEEE Inter. Conf. Acous., Spe., Sig. Proces.} IEEE, 2015, pp.
  16--20.

\bibitem{vantrees2002}
H.~L.~Van Trees,
\newblock {\em Optimum Array Processing (Detection, Estimation, and Modulation
  Theory, Part IV)},
\newblock John Wiley \& Sons, 2002.

\bibitem{music1986}
R.~Schmidt,
\newblock ``Multiple emitter location and signal parameter estimation,''
\newblock {\em IEEE Trans. Antennas Propag.}, vol. 34, no. 3, pp. 276--280,
  1986.

\bibitem{Compressivesensing}
D.~Malioutov, M.~{\c{C}}etin, and A.~S. Willsky,
\newblock ``A sparse signal reconstruction perspective for source localization
  with sensor arrays,''
\newblock {\em IEEE Trans. Signal Process}, vol. 53, no. 8, pp. 3010--3022,
  2005.

\bibitem{tipping2001}
M.~E. Tipping,
\newblock ``Sparse {B}ayesian learning and the relevance vector machine,''
\newblock {\em J. Mach. Learn. Res.}, vol. 1, pp. 211--244, Jun. 2001.

\bibitem{gerstoft2016}
P.~Gerstoft, C.~F. Mecklenbr{\"a}uker, A.~Xenaki, and S.~Nannuru,
\newblock ``Multisnapshot sparse {B}ayesian learning for {DOA},''
\newblock {\em IEEE Signal Process. Lett.}, vol. 23, no. 10, pp. 1469--1473,
  Oct. 2016.

\bibitem{pandey2021sparse}
R.~Pandey, S.~Nannuru, and A.~Siripuram,
\newblock ``Sparse {B}ayesian learning for acoustic source localization,''
\newblock in {\em IEEE Inter. Conf. Acous., Spe., Sig. Proces.} IEEE, 2021, pp.
  4670--4674.

\bibitem{kalman1960new}
R.~E. Kalman,
\newblock ``A new approach to linear filtering and prediction problems,''
\newblock {\em J. Basic Engineering}, vol. 82, no. 1, pp. 35--45, 1960.

\bibitem{pf2002tutorial}
M~Sanjeev Arulampalam, Simon Maskell, Neil Gordon, and Tim Clapp,
\newblock ``A tutorial on particle filters for online nonlinear/non-{G}aussian
  {B}ayesian tracking,''
\newblock {\em IEEE Trans. Signal Process.}, vol. 50, no. 2, pp. 174--188,
  2002.

\bibitem{park2021sequential}
Y.~Park, F.~Meyer, and P.~Gerstoft,
\newblock ``Sequential sparse {Bayesian} learning for time-varying direction of
  arrival,''
\newblock {\em J. Acoust. Soc. Am.}, vol. 149, no. 3, pp. 2089--2099, 2021.

\bibitem{opochinsky2021deep}
R.~Opochinsky, G.~Chechik, and S.~Gannot,
\newblock ``Deep ranking-based {DOA} tracking algorithm,''
\newblock in {\em 29th European Sig. Process. Conf. (EUSIPCO)}. IEEE, 2021, pp.
  1020--1024.

\bibitem{diaz2020robust}
D.~Diaz-Guerra, A.~Miguel, and J.~R. Beltran,
\newblock ``Robust sound source tracking using {SRP-PHAT} and 3{D}
  convolutional neural networks,''
\newblock {\em IEEE/ACM Trans. Audio, Speech, Language Process.}, vol. 29, pp.
  300--311, 2020.

\bibitem{wipf2007}
D.~P. Wipf and B.~D. Rao,
\newblock ``An empirical {B}ayesian strategy for solving the simultaneous
  sparse approximation problem,''
\newblock {\em IEEE Trans. Signal Process}, vol. 55, no. 7, pp. 3704--3716,
  2007.

\bibitem{compressivebeamforming}
A.~Xenaki, P.~Gerstoft, and K.~Mosegaard,
\newblock ``Compressive beamforming,''
\newblock {\em J. Acoust. Soc. Am.}, vol. 136, no. 1, pp. 260--271, 2014.

\bibitem{zhang2011sparse}
Z.~Zhang and B.~D. Rao,
\newblock ``Sparse signal recovery with temporally correlated source vectors
  using sparse {B}ayesian learning,''
\newblock {\em IEEE J. Sel. Topics Signal Process.}, vol. 5, no. 5, pp.
  912--926, 2011.

\bibitem{zhang2013extension}
Z.~Zhang and B.~D. Rao,
\newblock ``Extension of {SBL} algorithms for the recovery of block sparse
  signals with intra-block correlation,''
\newblock {\em IEEE Trans. Signal Process.}, vol. 61, no. 8, pp. 2009--2015,
  2013.

\bibitem{wipf2003bayesian}
D.~P. Wipf and B.~D. Rao,
\newblock ``Bayesian learning for sparse signal reconstruction,''
\newblock in {\em IEEE Inter. Conf. Acous., Spe., Sig. Proces.} IEEE, 2003,
  vol.~6, pp. VI--601.

\bibitem{SBLprespective}
J.~Palmer, B.~D. Rao, and D.~P. Wipf,
\newblock ``Perspectives on sparse {B}ayesian learning,''
\newblock in {\em Advances in neural info. proces. sys.}, 2004, pp. 249--256.

\bibitem{gemba2017multi}
K.~L. Gemba, S.~Nannuru, P.~Gerstoft, and W.~S. Hodgkiss,
\newblock ``Multi-frequency sparse {B}ayesian learning for robust matched field
  processing,''
\newblock {\em J. Acoust. Soc. Am.}, vol. 141, no. 5, pp. 3411--3420, 2017.

\bibitem{gemba2019robust}
K.~L. Gemba, S.~Nannuru, and P.~Gerstoft,
\newblock ``Robust ocean acoustic localization with sparse {B}ayesian
  learning,''
\newblock {\em IEEE J. Sel. Topics Signal Process.}, vol. 13, no. 1, pp.
  49--60, 2019.

\bibitem{multifreq2019sparse}
S.~Nannuru, K.~L. Gemba, P.~Gerstoft, W.~S. Hodgkiss, and C.~F.
  Mecklenbr{\"a}uker,
\newblock ``Sparse {B}ayesian learning with multiple dictionaries,''
\newblock {\em Signal Process.}, vol. 159, pp. 159--170, 2019.

\bibitem{nannuru2021sparse}
S.~Nannuru, P.~Gerstoft, G.~Ping, and E.~Fernandez-Grande,
\newblock ``Sparse planar arrays for azimuth and elevation using experimental
  data,''
\newblock {\em J. Acoust. Soc. Am.}, vol. 159, no. 1, pp. 167--178, 2021.

\bibitem{MLEefficient}
Z.M. Liu, Z.T. Huang, and Y.Y. Zhou,
\newblock ``An efficient maximum likelihood method for direction-of-arrival
  estimation via sparse {B}ayesian learning,''
\newblock {\em IEEE Trans. Wireless Commun.}, vol. 11, no. 10, pp. 1--11, 2012.

\bibitem{dempster1977maximum}
A.~P. Dempster, N.~M. Laird, and D.~B. Rubin,
\newblock ``Maximum likelihood from incomplete data via the {EM} algorithm,''
\newblock {\em J. Royal Statistical Society: Series B (Methodological)}, vol.
  39, no. 1, pp. 1--22, 1977.

\bibitem{bohme1985source}
J.F. Bohme,
\newblock ``Source-parameter estimation by approximate maximum likelihood and
  nonlinear regression,''
\newblock {\em IEEE J. Ocean. Eng.}, vol. 10, no. 3, pp. 206--212, 1985.

\end{thebibliography}

\end{document}